\documentclass[aps, twocolumn]{revtex4}
\usepackage{amssymb}
\usepackage{amsmath}
\usepackage{graphicx}
\usepackage{epsfig,amsmath}
\usepackage[bookmarksnumbered,linktocpage,pdfstartview=FitH]{hyperref}

\begin{document}

\title{Comment on "Dynamics of the Density of Quantized Vortex-Lines in
Superfluid Turbulence" }
\author{Sergey K. Nemirovskii\thanks{%
email address: nemir@itp.nsc.ru}}
\affiliation{Institute of Thermophysics, Lavrentyev ave, 1, 630090, Novosibirsk, Russia\\
and Novosibirsk State University, Novosibirsk}
\date{\today }

\begin{abstract}
In the paper by Khomenko et al. [Phys. Rev. B \textbf{91}, 180504 (2015)]
the authors, analyzing numerically the steady counterflowing helium in
inhomogeneous channel flow, concluded that the production term $\mathcal{P}$
in the Vinen equation is proportional to $\left\vert \mathbf{V}%
_{ns}\right\vert ^{3}\mathcal{L}^{1/2}$ (where $\mathcal{L}$ is vortex line
density and $\mathbf{V}_{ns}$ is the counterflow velocity). In present
comment we demonstrated that the procedure, implemented by the authors
includes a number of questionable steps, such as a decomposition of velocity
of line and interpretation of the flux term. Additionally, the overall
strategy - extracting information on the temporal behavior from the
stationary solution also remains questionable. Because of that the method of
determination of the explicit shape of Vinen equation is very sensitive to
the listed elements, the final conclusion of the authors cannot be
considered as unambiguous.
\end{abstract}

\maketitle

\section{Introduction.}

The question of dynamics of the vortex line density (VLD) $\mathcal{L}(%
\mathbf{r},t)$ of the vortex tangle (VT) is one of the most sacramental
problems in the theory of quantum turbulence. In fact the VLD is a crude
characteristic of flow, however it is responsible for many (mainly
hydrodynamic) phenomena in superfluid turbulence and knowing its exact
dynamics is very important for an adequate interpretation of various
experiments.

Long ago Vinen \cite{Vinen1957c} suggested that the rate of change of VLD $%
\partial {\mathcal{L}}(t)/\partial t$ can be described in terms of only
quantity ${\mathcal{L}}(t)$ itself (and also of other, external parameters,
such as counterflow velocity $\mathbf{V}_{ns}$ and temperature). He called
this statement as a self-preservation assumption. The according balance
equation for quantity ${\mathcal{L}}(\mathbf{r,}t)$ reads:
\begin{equation}
\frac{\partial \mathcal{L}(\mathbf{r},t)}{\partial t}=\mathcal{P}(\mathbf{r}%
,t)-\mathcal{D}(\mathbf{r},t).  \label{P3B}
\end{equation}%
Here $\mathcal{P}(\mathbf{r},t)$ is the so called production term, appeared
from interaction with external counterflow, $\mathcal{D}(\mathbf{r},t)$ is
the decay term. We imposed the spacial variable $\mathbf{r}$, having in mind
that we, after the authors of paper \cite{Khomenko2015}, will be discuss an
inhomogeneous channel flow. The form of the decay term Vinen extracted with
the use of some speculations. He concluded that $\mathcal{D}(\mathbf{r}%
,t)=C\kappa \mathcal{L}^{2}$, where $\kappa $ is the quantum of circulation,
and $C$ is a constant of the order of unity. Among different arguments, used
by Vinen, the dimensional analysis was undoubtedly main ingredient in the
whole procedure in getting quantity $\mathcal{D}(\mathbf{r},t)$.

The situation with the determination of the product term is much more
involved, from dimensional consideration it can be found only up to an
arbitrary function of a dimensionless argument $\kappa \sqrt{\mathcal{L}}/%
\mathbf{V}_{ns}$. The most accepted form of this term is $\mathcal{P}_{V}%
\mathcal{=}\alpha _{V}\left\vert \mathbf{V}_{ns}\right\vert \mathcal{L}%
^{3/2} $, which was obtained by Vinen \cite{Vinen1957c} on some speculations
and experimental measurements. Sometimes the so called alternative variant $%
\mathcal{P}_{alt}\mathcal{=}\alpha _{alt}\left\vert \mathbf{V}%
_{ns}\right\vert ^{2}\mathcal{L}$ is used. The problems of determination the
production term $\mathcal{P}$ were described in a review articles \cite%
{Nemirovskii1995},\cite{Nemirovskii2013}.

The authors of paper \cite{Khomenko2015} declared that "attempts to choose
the correct form of production term on the basis of time-dependent
homogeneous experiments ended inconclusively" and suggested their own way.
They\ tried to find the proper form of the production term studying the
steady case inhomogeneous flow. From their analysis Khomenko et al. have
concluded that the production term should have the form $\mathcal{P}_{3}%
\mathcal{=}\alpha _{3}\left\vert \mathbf{V}_{ns}\right\vert ^{3}\mathcal{L}%
^{1/2}$, the combination that has never been discussed earlier.

In this comment we demonstrate that this work includes a number of
questionable steps such as decomposition of velocity of line and
interpretation of the flux term. Since the method of determination of the
explicit form of the Vinen equation is very sensitive to the listed
elements, the final conclusion of the authors cannot be considered as
unambiguous.

\section{Questionable steps}

\subparagraph{Equation for the length of the vortex-line segment.}

The starting point for the evolution of the length $\delta \xi $ of the line
element $\mathbf{s(\xi )}$ is the relation
\begin{equation}
d\delta \xi /dt=(\mathbf{s}^{\prime }\cdot d\mathbf{\dot{s}}/d\xi )\delta
\xi .  \label{rate start}
\end{equation}%
This formula reflects the simple fact that the linear element changes its
length due to different velocities $\mathbf{\dot{s}}$ at the ends of
segment.\ The authors of paper \cite{Khomenko2015} used the following form
of this equation
\begin{equation}
\frac{1}{\delta \xi }\frac{d\delta \xi }{dt}=\alpha \mathbf{V}_{ns}(\mathbf{s%
},t)\cdot (\mathbf{s}^{\prime }\times \mathbf{s}^{\prime \prime }).
\label{ratesegment}
\end{equation}%
In this form, originally proposed by Schwarz \cite{Schwarz1978}, formula (%
\ref{ratesegment}) is valid only in the Local Induction Approximation.
Moreover, it is supposed that $\mathbf{V}_{ns}$ is spatially constant
(otherwise nonzero $d\mathbf{V}_{ns}(\mathbf{s},t)/d\xi $ changes the form
of Eq. (\ref{ratesegment})). It is important for discussed work, where the
full Biot-Savart law is applied.

\subparagraph{Velocity decomposition.}

The authors decomposed expression for $\mathbf{V}_{ns}$ (Eqs. (4b), (4c) of
paper \cite{Khomenko2015}) in the following manner
\begin{equation}
\mathbf{V}_{ns}=\mathbf{V}_{external}^{ns}-\mathbf{V}_{LIA}^{s}(\mathbf{s}%
,t)-\mathbf{V}_{nl}^{s}\ \ ,  \label{decompose}
\end{equation}%
where $\mathbf{V}_{external}^{ns}(\mathbf{s},t)$ is the external counterflow
velocity, created, e.g., by heat load, $\mathbf{V}_{LIA}^{s}(\mathbf{s},t)$
and $\mathbf{V}_{nl}^{s}(\mathbf{s},t)$ are the local and nonlocal parts of
superfluid velocity induced by the VT configuration. Then the authors used
the decomposition (\ref{decompose}) in the relation (\ref{ratesegment}) to
obtain an equation for the VLD evolution. At this stage the authors
associated the LIA part $\mathbf{V}_{LIA}^{s}(\mathbf{s},t)$ with the decay
term $\mathcal{D}(\mathbf{r},t)$, and nonlocal part $\mathbf{V}_{nl}^{s}(%
\mathbf{s},t)$ with the production term $\mathcal{P}(\mathbf{r},t)$. The
reason of this action is unclear. Indeed, in the Schwarz's formula (\ref%
{ratesegment}) the quantity $\mathbf{V}_{ns}$ is understood as an external
relative velocity, and reflects the fact that the VT length (and,
accordingly, its energy) grows due to external source. Eq. (6a) of paper
\cite{Khomenko2015} states that the production term works even when the $%
\mathbf{V}_{external}^{ns}$ is absent. This decisively contradicts to the
essence of the Feynman scenario of superfluid turbulence, and can be main
reason for new form of the Vinen equation Therefore the presence of nonlocal
quantity $\mathbf{V}_{nl}^{s}(\mathbf{s},t)$ in the production term looks
unmotivated. At the same time I agree that nonlocal velocity results in the
growth of the VLD, but this effect should be evaluated in a different
manner. It should be accounted by the use of starting Eq. (\ref{rate start}%
). In this case the nonlocal part results in the stretching of lines (even
when the mutual friction is absent, $\alpha =0$), unlike the local part.

\subparagraph{Vortex-line density flux.}

To take into account inhomogeneous flows, the author introduced the \
vortex-line density flux\emph{\ }$\nabla \cdot \mathbf{J}(\mathbf{r},t)$
into the balance equation (see Eq. (3b) of paper \cite{Khomenko2015}). The
authors associated it with the drift motion of the VT $\mathbf{V}_{drift}$.
It is defined in the middle part of Eq. (6.c) of paper \cite{Khomenko2015})
and coincides with definition proposed by Schwarz (see \cite{Schwarz1988},
Eq. (21)). However in Schwarz's paper the drift motion is associated with
the full velocity $\dot{\mathbf{s}}$, whereas the choice of $\mathbf{V}%
_{drift}$ in paper \cite{Khomenko2015} (right hand side of Eq. (6c) and
equation (7) in their work) is not complete, the self-induced motion is
missed. But this, missed self-induced motion generates "irreversible" flux
of VLD $\mathcal{L}(\mathbf{r},t)$, realized by emission of vortex loops,
(see, e.g.,\cite{Barenghi2002},\cite{Nemirovskii2010},\cite{Kondaurova2012}).

Furthermore, it is obviously that flux of quantity $\mathcal{L}$ must have
the structure of sort $\mathcal{L}\mathbf{V}$. As a matter of fact this
structure presents in latent form in Eq. (6c) of paper \cite{Khomenko2015},
where VLD came from integration over $d\xi $. But later the closure version
for flux ( Eq. (10)) was taken that it does not include VLD $\mathcal{L}$ at
all. It is not motivated and looks strange.

Summarizing this part of our comments we would like to emphasize, that the
adjustment of curves (6a)-(6c), extracted from numerical simulations, to the
their closure counterparts (3a-3c, 1c, 3c) crucially is dependent on
questionable steps, described above. Therefore the conclusion about new form
of production term (3a) cannot be considered as definitely proven.

\section{ Temporal vs Spacial}

The second point concerns the principal idea of work \cite{Khomenko2015},
which the authors have stated in the abstract of their paper as: "To
overcome this difficulty we announce here an approach that employs an
(steady) inhomogeneous channel flow which is excellently suitable to
distinguish the implications of the various possible forms of the desired
equation." The applicability of such approach seems to be also questionable.
Let me give the nearest and simplest counterexample. Take, for instance some
quantity $\Psi (\mathbf{r,}t).$ Consider three different equation
\begin{gather}
\dot{\Psi}(\mathbf{r,}t)=\mathcal{\hat{F}}\{\Psi (\mathbf{r,}t)\},\ \ \dot{%
\Psi}(\mathbf{r,}t)=K\mathcal{\hat{F}}\{\Psi (\mathbf{r,}t)\},\ \
\label{t vs r} \\
\text{and \ \ }\dot{\Psi}(\mathbf{r,}t)=(\mathcal{\hat{F}}\{\Psi (\mathbf{r,}%
t)\})^{2},  \notag
\end{gather}%
where $\mathcal{\hat{F}}$ is some operator acting on spatial variables, and $%
K$ is arbitrary number. In the steady case all three equations produce the
same function $\Psi _{eq}(\mathbf{r,}t)$, which satisfies equation $\mathcal{%
\hat{F}}\{\Psi _{eq}(\mathbf{r,}t)\}=0.$ At the same time, in the
nonstationary situation the temporal evolution of all quantities $\Psi (%
\mathbf{r,}t)$ (which are not equal to $\Psi _{eq}(\mathbf{r,}t)$ ) will be
absolutely different. This counterexample convinces that the way, chosen by
the authors of paper \cite{Khomenko2015} is not applicable. Or, any
additional argumentation must be presented.

\section{ Stationary solution}

In fact, problems arise already at a stage of stationary solution. The point
is that the equation for vortex line density derived in paper \cite%
{Khomenko2015} and results on the distribution of quantity $\mathcal{L(}y)$
are inconsistent. Indeed, Let's take the stationary variant of equation for
VLD offered in paper \cite{Khomenko2015}, (Eq. (3b)) with the "closure"
terms (1c), (3a) (3c) and the coefficients $C_{flux}$ , $C_{prod}$ , $%
C_{dec} $, given in Table I.
\begin{equation}
-\frac{\alpha }{2\kappa }C_{flux}\frac{\partial ^{2}\mathbf{V}_{ns}^{2}}{%
\partial y^{2}}=\frac{\alpha C_{prod}}{\kappa ^{2}}\left\vert \mathbf{V}%
_{ns}(y)\right\vert ^{3}\mathcal{L}^{1/2}-\alpha C_{dec}\kappa \mathcal{L}%
^{2}.  \label{balance}
\end{equation}%
Let's take further the velocity $\mathbf{V}_{ns}(y)$ profile from Fig. 1 of
paper \cite{Khomenko2015}. It is seen that with the good accuracy it can
extrapolated by usual parabolic shape $(\rho /\rho _{s}\mathbf{V}%
_{n}(y))^{2} $. The final equation is usual algebraic equation of forth
order with respect to variable $\mathcal{L}^{1/2}$. \ The according solution
leads to that with the good precision $\mathcal{L(}y)\approx (\alpha
C_{prod}/C_{dec}\kappa ^{3})^{2/3}\left\vert \mathbf{V}_{ns}(y)\right\vert
^{2}$, which is very far from $\mathcal{L(}y)$, proposed in paper \cite%
{Khomenko2015} (Fig. 1, Panels b). This inconsistency demonstrated that
either the choice of the closure form is not correct, or something is missed.

\section{VINEN EQUATION}

The third issue, that I would like to discuss concerns the Vinen-type
equation itself and its possible variants. In general, the Vinen equation
(in any variant ((2a),(2b),(3a)) of production term) is not valid. Indeed,
assume that one changes velocity $\mathbf{V}_{ns}(\mathbf{s},t)$
instantaneously on the opposite. Since all the listed forms of the
Vinen-type equation include the absolute value of relative velocity $%
\left\vert \mathbf{V}_{ns}(\mathbf{s},t)\right\vert $, then, from a formal
point of view, nothing will happen. That is wrong, of course. The structure
of the VT, mean curvature, anisotropy and polarization parameters will
become reorganized. That implies the violation of the self-preservation
assumption, and dynamics of the VLD ${\mathcal{L}}(t)$ depends on other,
more subtle characteristics of the vortex structure, different from ${%
\mathcal{L}}(t)$.

Meantime, it is intuitively seems truthful, that for slow changes (both in
space and in time) the self-preservation assumption takes place. Some
justification of this assertion is that the resulting model describes well
the experimental observations on the strong heat pulses propagation, which
generate quantum vortices and interact with them ( see review article \cite%
{Nemirovskii1995}). It is understood that the equation $\partial {\mathcal{L}%
}(t)/\partial t=\mathcal{F}({\mathcal{L}})$ can be used unless we are not
interested in the special problems related to the fine structure of the VT.

To clarify situation, let's consider a way of derivation of VE from the
dynamics of vortex filaments in the local induction approximation. It is
enough for illustration. Integrating Eq. (\ref{ratesegment}) over $\xi $ in
volume $\Omega $ \ we have, that in the counterflowing helium II VLD ${%
\mathcal{L}}(t)$ obeys equation \cite{Schwarz1978}, \cite{Nemirovskii2007a})
(cf. with Eqs. (6a), (6b) in \cite{Khomenko2015})
\begin{equation}
\frac{\partial \mathcal{L}}{\partial t}=\;\frac{\alpha \mathbf{V}_{ns}}{%
\Omega }\int \left\langle \mathbf{s}^{\prime }\times \mathbf{s}^{\prime
\prime }\right\rangle \;d\xi \;-\frac{{\alpha \beta }}{\Omega }\int
\left\langle |\mathbf{s}^{\prime \prime }|^{2}\right\rangle \;d\xi \;.
\label{VLD rate}
\end{equation}%
Notation are in \cite{Khomenko2015}. Quantity ${\mathcal{L}}(t)$ is related
to the first derivative of function $\mathbf{s}^{\prime }$ (${\mathcal{L}}%
(t)\propto \int |\mathbf{s}^{\prime }|d\xi \;$). The rate of change of ${%
\mathcal{L}}(t)$ includes quantities with a higher-order derivative $\mathbf{%
s}^{\prime \prime }$, namely $\left\langle \mathbf{s}^{\prime }\times
\mathbf{s}^{\prime \prime }\right\rangle $ and $\left\langle |\mathbf{s}%
^{\prime \prime }|^{2}\right\rangle $. In steady these quantities are
related to VLD $\mathcal{L}$ as $\left\langle \mathbf{s}^{\prime }\times
\mathbf{s}^{\prime \prime }\right\rangle \propto $ $I_{l}\mathcal{L}^{1/2%
\text{ }}$and $\left\langle |\mathbf{s}^{\prime \prime }|^{2}\right\rangle
\propto c_{2}^{2}(T)\mathcal{L}$ ($I_{l},c_{2}(T)$ is a temperature
dependent parameters, introduced by Schwarz \cite{Schwarz1988}). But in the
nonstationary situation $\mathbf{s}^{\prime \prime }$ is a new independent
variables, and one needs the new independent equation for it and for other
quantities, related to curvature. This new equation, in turn, includes
higher derivatives $\mathbf{s}^{\prime \prime \prime },\mathbf{s}^{IV}$ and
so on. This infinite chain can be cut if, for some reasons, the higher-order
derivatives relax faster, than the low-order derivatives, and take their
"equilibrium" values (with respect to the moments of low order). Applying
this speculations to equation (\ref{VLD rate}) we are arrive at the the
following bifurcation:

1. Time of relaxation of VLD $\mathcal{L}(t)$ is much larger than that for
quantities with higher derivatives, and the latter have enough time to
adjust to change of $\mathcal{L}(t)$, i.e. $\left\langle \mathbf{s}^{\prime
}\times \mathbf{s}^{\prime \prime }\right\rangle \propto $ $I_{l}\mathcal{L}%
^{1/2\text{ }}$. Then, the self-preservation assumption\ is valid, and
generating term has a classical form $\mathcal{P}_{V}\mathcal{=}\alpha
_{V}\left\vert \mathbf{V}_{ns}\right\vert \mathcal{L}^{3/2}$.

2 Time of relaxation of VLD $\mathcal{L}(t)$ is of the same order as that
for quantities with higher derivatives. Then, the self-preservation
assumption\ is not valid, substitution of $\left\langle \mathbf{s}^{\prime
}\times \mathbf{s}^{\prime \prime }\right\rangle \propto $ $I_{l}\mathcal{L}%
^{1/2\text{ }}$ is inadmissible, and, there is no theoretical grounds to cut
a chain. Thus, in general, no equation of type{\Large \ }$\partial {\mathcal{%
L}}(t)/\partial t=\mathcal{F}({\mathcal{L}},\mathbf{V}_{external}^{ns}(t))$%
{\Large \ }exists! \ At the same time under some (unclear) conditions, and
with the use of additional arguments (see, \cite{Vinen1957c}) it is possible
to write down the required equation. However in this case, the region of
applicability of this equation is not clear\emph{, }see my example above
with the sudden inversion of the counterflow velocity. Therefore, Vinen
equation in it classical form should be considered as a good approximation
for applied, engineering problems (for instance, in study of propagation of
large thermal pulses).

\section{Conclusion}

Resuming, I would state \emph{that the final, sensational conclusion of the
authors of \cite{Khomenko2015}, asserting that the production term} $%
\mathcal{P}_{3}\mathcal{=}\alpha _{3}\left\vert \mathbf{V}_{ns}\right\vert
^{3}\mathcal{L}^{1/2}$, \emph{cannot be considered as unambiguous}.
Undoubtedly, the authors raised interesting and actual question of
inhomogenious quantum turbulence, but the change the form of the Vinen
equation seems premature.

I would like to thank the authors of paper \cite{Khomenko2015}\emph{\ }D.
Khomenko, L. Kondaurova, V. L'vov, P. Mishra, A. Pomyalov, and I. Procaccia
for very fruitful discussion on macroscopic quantum turbulence.

The work was supported by Grant No. 14-19-00352 from RSCF (Russian
Scientific Foundation)


\end{document}